\documentclass{article}
\usepackage{amsmath,amssymb,amstext,mathtools,array,url,bm,graphicx,color,epsfig}
\usepackage{fullpage,setspace}
\usepackage{authblk}
\usepackage{filecontents}
\usepackage{natbib}
\usepackage{lineno}
\usepackage{subcaption}
\usepackage{float}
\usepackage[flushleft]{threeparttable}

\begin{document}
\title{Notes on the analytic solution of box model equations for gravity-driven particle currents with constant volume.}
\author[]{\small Andrea Bevilacqua}
\affil[]{\textit{Sezione di Pisa, Istituto Nazionale di Geofisica e Vulcanologia, Pisa, Italy}}
\date{}

\maketitle
\abstract
We summarize the physical equations and analytic solutions of three versions of the box model equations, suitable for the integral formulation of axisymmetric gravity-driven particle currents with constant volume. The first model is based on a simple constant resisting stress, while the second and third models assume flow dilution by particle deposition. The third model is characterized by assuming an interstitial fluid lighter than the ambient fluid. All the calculations are performed on a flat topography. Ambient fluid entrainment and cooling effects are not considered. All particles are assumed to deposit at the same velocity.

\section{Introduction}
The box model integral formulation for gravity-driven particle currents is based on the pioneering work of \cite{HuppertSimpson1980}. The theory is detailed in \cite{Bonnecaze1995, Hallworth1998}. We assume axisymmetric geometry and constant volume of the flow\footnote{this is not the formulation adopted in \cite{DadeHuppert1996}, which assumed constant volume flux.}. We solve the equations over a flat topography, with no slope or obstacles. Ambient fluid entrainment is neglected and thermal properties of the flow remain constant\footnote{air entrainment and cooling effects are explored in \cite{BursikWoods1996, Fauria2016}.}. In the second and third model all the solid particles are assumed to deposit at the same velocity, thus neglecting differences in particle size and shape.

\section{Box model with constant resisting stress}
This model is described in \cite{DadeHuppert1998} and it is at the base of the depth-averaged model in \cite{Kelfoun2009}. We consider the work $W(t)$ done by a constant resisting stress $\tau_c$ acting over the basal area $A$ of the flow. We assume the flow geometry to be cylindrical, so $A(r)=\lambda r^2$, where $r$ is the cylinder radius.
\\
Let us consider the time interval $[0,t]$ and assume $L(0)=0$. We obtain:
$$W(t)=\int_0^{L(t)} \tau_c \cdot A(r) dr=\int_0^{L(t)} \tau_c \cdot \lambda r^2 dr=\frac{\tau_c}{3} \lambda L(t)^3=\tau \lambda L(t)^3,$$
where $L(t)$ is the radial distance reached by the current at time $t$, $\lambda$ is half-central angle of the cylindrical sector considered, and $\tau:=\frac{\tau_c}{3}$ is the effective stress constant.
\\
Let us also assume that the total energy of the flow at $t_0=0$ is:
$$Q(t_0)=g H \rho V,$$
where $g$ is gravity, $\rho$ is the density of the flowing material, $H$ is the height drop of the flow, and $V$ its volume. We finally assume that the volume $V$ of the current is preserved.
\\
Hence, the kinetic energy of the flow at time $t$ is given by:
$$K(t)=g H \rho V - \lambda\tau L(t)^3.$$
Assuming $K(t_f)=0$ at the instant $t_f$ at which the runout distance $L(t_f)$ is reached, we obtain:
$$V=\frac{\lambda \tau L(t_f)^3}{g H \rho}.$$

\section{Box model with particle deposition}
This model is described in \cite{DadeHuppert1995} and it adopted in \cite{Neri2015, Bevilacqua2017}. Further description of this approach is provided in \cite{Bevilacqua2016, EOngaro2016}. We consider the Von K\'arm\'an equation for density currents:
$$\frac{dL}{dt}(t)=Fr \sqrt{h(t) g\frac{\rho_c(t)-\rho_a}{\rho_a}},$$
where $L(t)$ and $h(t)$ are the radial distance reached by the current and its height, at time $t$. In our notation $Fr$ is the Froude Number, $g$ is gravity. We express the density of the current $\rho_c$ by:
$$\rho_c(t)=\phi(t)\rho + \left[1-\phi(t)\right]\rho_a$$
where $\rho$ is the density of the solid particles, and $\rho_a$ is the density of ambient fluid. Finally, the equation
$$\frac{d\phi}{dt}(t)=-w_s \frac{\phi(t)}{h(t)}$$
defines the particle volume fraction $\phi(t)$ at time $t$.
\\
Let us also assume that the volume $V$ of the current is preserved, and that its geometry is cylindrical:
$$V=L(t)^2 h(t) \lambda = const,$$
where $\lambda$ is half-central angle of the cylindrical sector considered ($\lambda=\pi$ in axisymmetric examples).
\\
In summary:
$$\left\{
  \begin{array}{ll}
    \frac{dL}{dt}(t)=Fr \sqrt{h(t) \phi(t) g\frac{\rho-\rho_a}{\rho_a}}, \\
\\
    \frac{d\phi}{dt}(t)=-w_s \frac{\phi(t)}{h(t)}, \\
\\
    V\equiv L(t)^2 h(t) \lambda.
  \end{array}
\right.$$

\subsection{Derivation of the equation for the volume V}
Let us consider the formal expression:
$$\frac{d\phi}{dL}(t)=\frac{d\phi}{dt}(t)\left[\frac{dL}{dt}(t)\right]^{-1},$$
and so
$$\frac{d\phi}{dL}(t)=-w_s \frac{\phi(t)}{h(t)}\left[Fr \sqrt{h(t) \phi(t) g_p}\right]^{-1},$$
where we use the short notation $g_p:= g\frac{\rho-\rho_a}{\rho_a}$.
\\
If we plug in the expression $h(t)=\frac{V}{\lambda L(t)^2}$, we obtain:
$$\frac{d\phi}{dL}(t)=-w_s \phi^{1/2}(t)L(t)^3\left[Fr\ g_p^{1/2} \frac{V}{\lambda}^{3/2}\right]^{-1}.$$
\\
If we define
$$\eta:=w_s \left(Fr\ g_p^{1/2} \frac{V}{\lambda}^{3/2}\right)^{-1},$$
we can write:
$$\phi(t)^{-1/2}d\phi=-\eta L^3 dL.$$
\\
If we integrate the differential expression over $[0,t]$, assuming $\phi(0)=\phi_0$ and $L(0)=0$, we get:
$$\phi^{1/2}(t)=\phi_0^{1/2} -\frac{1}{8}\eta L(t)^4.$$
\\
Finally, assuming $\phi(t_f)=0$ at the instant $t_f$ at which the runout distance $L(t_f)$ is reached, we obtain:
$$L(t_f)=\left(\frac{8\phi_0^{1/2}}{\eta}\right)^{1/4}=\left[8\frac{Fr\ \phi_0^{1/2} g_p^{1/2} (V/\lambda)^{3/2}}{w_s}\right]^{1/4},$$
and so
$$V=\lambda \left(\frac{L^4 w_s}{8\phi_0^{1/2}g_p^{1/2}Fr}\right)^{2/3}=\lambda \frac{L^{8/3} w_s^{2/3}}{4\phi_0^{1/3}g_p^{1/3}Fr^{2/3}}.$$

\section{Two-phase box model (interstitial fluid + solid particles)}
We consider the Von K\'arm\'an equation for density currents:
$$\frac{dL}{dt}(t)=Fr \sqrt{h(t) g\frac{\rho_c(t)-\rho_a}{\rho_a}},$$
where $L(t)$ and $h(t)$ are the radial distance reached by the current and its height, at time $t$. In our notation $Fr$ is the Froude Number, $g$ is gravity. We express the density of the current $\rho_c$ by:
$$\rho_c(t)=\phi(t)\rho + \left[1-\phi(t)\right]\rho_i$$
where $\rho$ is the density of the solid particles, and $\rho_i$ is the density of interstitial fluid. We have that:
$$\frac{\rho_c-\rho_a}{\rho_a}=\frac{\phi(t)\rho + \left[1-\phi(t)\right]\rho_i-\rho_a}{\rho_a}=\phi(t)\frac{\rho - \rho_i}{\rho_a}+\frac{\rho_i - \rho_a}{\rho_a}=\left[\phi(t)-\phi_{cr}\right] \frac{\rho-\rho_i}{\rho_a},$$
where we called $\phi_{cr}:=\frac{\rho_a-\rho_i}{\rho - \rho_i}$.
\\
Finally, the equation
$$\frac{d\phi}{dt}(t)=-w_s \frac{\phi(t)}{h(t)}$$
defines the particle volume fraction $\phi(t)$ at time $t$.
\\
Let us also assume that the volume $V$ of the current is preserved, and that its geometry is cylindrical\footnote{the analytic solutions in cartesian geometry are provided in \cite{EOngaro2016}.}:
$$V=L(t)^2 h(t) \lambda = const,$$
where $\lambda$ is half-central angle of the cylindrical sector considered ($\lambda=\pi$ in axisymmetric examples).
\\
In summary:
$$\left\{
  \begin{array}{ll}
    \frac{dL}{dt}(t)=Fr \sqrt{h(t) \left[\phi(t) - \phi_{cr}\right] g\frac{\rho-\rho_i}{\rho_a}}, \\
\\
    \frac{d\phi}{dt}(t)=-w_s \frac{\phi(t)}{h(t)}, \\
\\
    V\equiv L(t)^2 h(t) \lambda.
  \end{array}
\right.$$

\subsection{Derivation of the equation for the volume V}
Let us consider the formal expression:
$$\frac{d\phi}{dL}(t)=\frac{d\phi}{dt}(t)\left[\frac{dL}{dt}(t)\right]^{-1},$$
and so
$$\frac{d\phi}{dL}(t)=-w_s \frac{\phi(t)}{h(t)}\left[Fr \sqrt{h(t) \left[\phi(t) - \phi_{cr}\right] g_c}\right]^{-1},$$
where we use the short notation $g_c:=g\frac{\rho-\rho_i}{\rho_a}$.
\\
If we plug in the expression $h(t)=\frac{V}{\lambda L(t)^2}$, we obtain:
$$\frac{d\phi}{dL}(t)=-w_s \frac{\phi(t)}{\sqrt{\phi(t)-\phi_{cr}}}L(t)^3\left[Fr\ g_c^{1/2} \frac{V}{\lambda}^{3/2}\right]^{-1}.$$
\\
If we define the function:
$$\hat\phi(t)=\frac{\phi(t)}{\phi_{cr}},$$
and the constant
$$\eta:=w_s \left(Fr\ g_c^{1/2} \frac{V}{\lambda}^{3/2}\right)^{-1},$$
we can write:
$$\frac{d\hat\phi}{dL}(t)=-\eta \frac{\hat\phi(t)}{\phi_{cr}^{1/2}\left(\hat\phi(t)-1\right)^{1/2}},$$
and then
$$\frac{\phi_{cr}^{1/2}\left(\hat\phi(t)-1\right)^{1/2}}{\hat\phi(t)} d\hat\phi=-\eta L^3 dL.$$
\\
If we integrate the differential expression over $[0,t]$, assuming $\hat\phi(0)=\hat\phi_0$ and $L(0)=0$, we get:
$$F\left(\hat\phi(t)\right) = F\left(\hat\phi_0\right) -\frac{\eta}{4\phi_{cr}^{1/2}} L(t)^4,$$
where we defined the function:
$$F(x):= 2\left[\sqrt{x-1} - \arctan\left(\sqrt{x-1}\right)\right].$$
Indeed:
$$\int \frac{\sqrt{x-1}}{x} dx = \int \frac{2z^2}{z^2+1} dz = 2\int dz - 2\int \frac{1}{z^2+1} dz=2\left[z-\arctan(z)\right] + c,$$
where $z:=\sqrt{x-1}$.
\vskip.3cm\noindent
Finally, assuming $\hat\phi(t_f)=1$ at the instant $t_f$ at which the runout distance $L(t_f)$ is reached, we obtain:
$$L(t_f)=\left(\frac{4\phi_{cr}^{1/2}\ F\left(\hat\phi_0\right)}{\eta}\right)^{1/4}=\left[\frac{8\phi_{cr}^{1/2}\left[\sqrt{\frac{\phi_0}{\phi_{cr}}-1}-\arctan\left(\sqrt{\frac{\phi_0}{\phi_{cr}}-1}\right)  \right] g_c^{1/2} Fr\ (V/\lambda)^{3/2}}{w_s}\right]^{1/4},$$
and so
$$V=\lambda \left(\frac{L^4 w_s}{4\phi_{cr}^{1/2}\ F\left(\hat\phi_0\right)\ g_c\ Fr}\right)^{2/3}=\lambda \frac{L^{8/3} w_s^{2/3}}{4\phi_{cr}^{1/3} \left[\sqrt{\frac{\phi_0}{\phi_{cr}}-1}-\arctan\left(\sqrt{\frac{\phi_0}{\phi_{cr}}-1}\right)\right]^{2/3}g_c^{1/3}Fr^{2/3}}.$$

\bibliographystyle{apalike}
\bibliography{bibfile}
\end{document}